\RequirePackage{fix-cm}
\documentclass[twocolumn,epjc3]{svjour3}  
\smartqed  % flush right qed marks, e.g. at end of proof
\RequirePackage{graphicx}
\RequirePackage{float}
\RequirePackage{hyperref}
\RequirePackage{amsmath}
\usepackage{mwe}
\RequirePackage{amssymb}
\RequirePackage{mathtools}   
\usepackage{orcidlink}
\usepackage{booktabs}
\usepackage{caption}
\usepackage{multirow}
\usepackage{dcolumn}
\usepackage[caption=false]{subfig}
\usepackage{float}   % use Times fonts if available on your TeX system
\journalname{Eur. Phys. J. C}
\begin{document}

\title{Viscous cosmology in the Weyl-type $f(Q,T)$ gravity}
%\subtitle{Do you have a subtitle?\\ If so, write it here}

%\titlerunning{Short form of title}        % if too long for running head

\author{Gaurav N. Gadbail\thanksref{e1,addr1}
        \and
        Simran Arora\thanksref{e2,addr1}
        \and
        P.K. Sahoo\thanksref{e3,addr1}}

%\thankstext{t1}{Grants or other notes
%about the article that should go on the front page should be
%placed here. General acknowledgments should be placed at the end of the article.
\thankstext{e1}{e-mail: gauravgadbail6@gmail.com}
\thankstext{e2}{e-mail: dawrasimran27@gmail.com}
\thankstext{e3}{e-mail: pksahoo@hyderabad.bits-pilani.ac.in}

%\authorrunning{Short form of author list} % if too long for running head

\institute{Department of Mathematics, Birla Institute of
Technology and Science-Pilani, Hyderabad Campus, Hyderabad-500078,
India \label{addr1}
          }

\date{Received: date / Accepted: date}
% The correct dates will be entered by the editor

\maketitle

\begin{abstract}
Bulk viscosity is the only viscous influence that can change the background dynamics in a homogeneous and isotropic universe. In the present work, we analyze the bulk viscous cosmological model with the bulk viscosity coefficient of the form $\zeta=\zeta_0+\zeta_1H+\zeta_2\left(\frac{\dot{H}}{H}+H\right)$ where, $\zeta_0$, $\zeta_1$ and $\zeta_2$ are bulk viscous parameters, and $H$ is the Hubble parameter. We investigate the impact of the bulk viscous parameter on dynamics of the universe in the recently proposed Weyl-type $f(Q, T)$ gravity, where $Q$ is the non-metricity, and $T$ is the trace of the matter energy-momentum tensor. The exact solutions to the corresponding field equations are obtained with the viscous fluid and the linear gravity model of the form $f(Q, T)=\alpha Q+\frac{\beta}{6\kappa^2}T$, where $\alpha$ and $\beta$ are model parameters. Further, we constrain the model parameters using the $57$ points Hubble dataset and recently released 1048 points Pantheon sample and the combination Hz+BAO+Pantheon, which shows our model is good congeniality with observations. We study the possible scenarios and the evolution of the universe through the deceleration parameter, the equation of state (EoS) parameter, the statefinder diagnostics, and the Om diagnostics. It is observed that the universe exhibits a transition from a decelerated to an accelerated phase of the universe under certain constraints of model parameters.
% \subclass{MSC code1 \and MSC code2 \and more}
\end{abstract}

\section{Introduction}\label{sec1}

The studies of high redshift supernovae \cite{Perlmutter/1999,Riess/1998,Riess/2004}, WMAP data \cite{Hanany/2000,Spergel/2007}, Cosmic Microwave Background (CMB) peaks \cite{Komatsu/2011}, Baryon Acoustic Oscillations (BAO) \cite{Daniel/2005} have provided fascinating evidence that the universe contains $95\%$ of its contents in two unknown forms of energy and matter, which we refer to as dark energy (DE) with negative pressure and dark matter (DM). These observations and data conclude that our universe is not only expanding but also accelerating. Finding the responsible candidate for this accelerating expansion is the most fundamental question in cosmology. There are a variety of proposed models for understanding the accelerated expansion. The most well-known is Einstein's equations being modified to include a cosmological constant $\Lambda$. Despite being extremely successful, this method has significant issues like the cosmic coincidence problem and fine-tuning problem. As a result, this leads to investigating alternative possibilities such as dynamical models   (modifying matter content): Chapylygin gas, Quintessence, k-essence, etc. Recently, to understand the DE problem, the modified theory of gravity has become one of the most admired candidates. Many authors have discussed several modified theories of gravity to explain the early and late-time acceleration of the universe. 
Some of the modified theories include the $f(R)$ theory of gravity, the modification of GR, introduced in \cite{Buchdahl/1970}, the $f(R,T)$ theory, an extension of $f(R)$ gravity coupled with the energy-momentum tensor $T$ \cite{Harko/2011,Shabani/2013,Sahoo/2018}, the $f(G)$ gravity \cite{Felice/2009,Goheer/2009,Bamba/2017}, $f(R,G)$ theory \cite{Elizalde/2010,Bamba/2010}, and others. \\
Commonly, the most generic affine connection can be split into three components: Christoffel symbol, contorsion tensor, and deformation tensor-relates to non-metricity. So, both torsion and non-metricity vanish in GR, resulting in the Levi-Civita connection. However, non-zero torsion and non-metricity leads to alternative theories of gravity such as the teleparallel or $f(T)$ gravity, where $T$ is torsion \cite{Ferraro/2007,Myrzakulov/2011,Capozziello/2011}, and the symmetric teleparallel or $f(Q)$ gravity, where $Q$ is the non-metricity \cite{Jimenez/2018}. Yixin et al. \cite{Xu/2019} recently investigated the $f(Q, T)$ theory of gravity as a, extension of symmetric teleparallel gravity or $f(Q)$ gravity, which is based on nonminimal coupling between the non-metricity $Q$ and trace of the matter energy-momentum tensor $T$. They analyzed the cosmological implication for three types of specific models in the $f(Q,T)$ theory. Their outcomes, with the solution, described both the accelerating and decelerating evolutionary phase of the universe. Numerous studies have shown that $f(Q, T)$ gravity is a viable approach to explaining current cosmic acceleration and providing a consistent solution to the dark energy problem \cite{Arora/2020,Arora/2021}.\\
In the present article, we will investigate the recently proposed Weyl type $f(Q, T)$ gravity. Weyl gravity generalizes GR by admitting non-metricity in the affine connection. The original form of the proposed Weyl gravity unifies gravity and electromagnetism. Recently, the Weyl gravity has been revived to solve the dark matter and dark energy problems or inflation \cite{Alvarez/2017}. Gomes et al. \cite{Gomes/2019} investigated the non-minimal coupling between the curvature and matter in the Weyl gravity theory. In the framework of proper Weyl geometry, Yixin et al.  \cite{Xu/2020} introduced a non-minimal coupling between non-metricity $Q$ and the trace of energy-momentum tensor, where the non-metricity is completely determined by the magnitude of the vector field $\omega_{\mu}$. In the Weyl geometry, the general field equations of the model, obtained from a variation of action with respect of metric tensor, allow a complete description of the gravitational phenomena in terms of a vector field $w_{\mu}$. Despite its newness, the Weyl type $f(Q, T)$ gravity theory has a numerous intriguing and valuable applications in the literature. The accelerating and decelerating evolutionary phases of the universe, the Newtonian limit, geodesic and Raychaudhuri equation, tidal forces, power-law solution in the Weyl type $f(Q, T)$ gravity are listed in References \cite{Xu/2020,Yang/2021,Gadbail/2021,Wheeler/2014}.\\
According to current understanding, the expansion of the universe is a result of negative pressure in the cosmic fluid. This has led to renewed interest in treating the fluid as a two-component mixture: a standard fluid component and a dark energy component \cite{Brevik/2012}. A significant role is played by dissipative processes such as heat transport, bulk viscosity,  and shear viscosity in cosmic expansion. Moreover, one can omit shear viscosity in the homogeneous and isotropic FLRW background. Concern over the bulk viscosity is appropriate at this moment as we do not know the nature of the content of the universe: dark energy and dark matter components. Relaxation mechanisms related to bulk viscosity efficiently reduce pressure in an expanding system, making it negative and perhaps mimicking dark energy behavior. Several authors have investigated the effect of bulk viscous content in different cosmological contexts like viscous dark energy \cite{Capozziello/2006,Colistete/2007,Feng/2009}, viscous dark matter \cite{Zimdahl/2001,Velten/2012}, inflation in a viscous fluid model \cite{Bamba/2016}, late-time cosmic acceleration \cite{Solanki/2021,Arora/2021}, etc. Brevik et al. \cite{Brevik/2020,Brevik/2013} established the critical role of viscosity in explaining the inflationary era and the  current epoch of the universe. Later, Padmanabhan and Chitre \cite{Pad/1987} used the concept of bulk viscosity to characterize inflation in the early universe. Additionally, the relationship between the modified EoS of viscous cosmological models, scalar fields, and extended theories of gravity is being examined \cite{SCap/2006,Ren/2006}. Deng et al. \cite{Deng/2017} proposed innovative viscous dark energy models that were constrained by current cosmic observations. Mark and Harko \cite{Mark/2003} looked at the effect of bulk viscosity in the Brans-Dicke gravitational theory. Cataldo et al. \cite{Cataldo/2005} established a bulk viscous cosmological solution at the boundary with a big rip singularity. The role of bulk viscosity with the linear viscous coefficient in modified $f(R, T)$ theory has been studied by Singh and Kumar \cite{Singh/2014}. Davood \cite{Davood/2019} also examined the influence of bulk viscous matter in $f(T)$ gravity.  Srivastava and Singh \cite{Srivastava/2018} studied new holographic dark energy (HDE) model in modified $f(R, T)$ gravity theory within the framework of an FLRW model with bulk viscous matter content. Some works in other gravity theories can be checked  in \cite{Mahanta/2014,Sharif/2013,Sharif/2017}.  The $f(Q,T)$ gravity is another non-minimally coupled model for correcting the gravity sector. Hence, we intend to study the cosmic acceleration under the impact of bulk viscosity in the framework of the Weyl type $f(Q, T)$ gravity.\\ 
The majority of research has focused on observable evidence from type Ia supernovae, the cosmic microwave background, baryon acoustic oscillations, and the Hubble estimates, all of which contribute to the constraining of cosmological models \cite{Jimenez/2002}. So, we test the model by using recent observational data which includes the estimates of the Hubble parameter $H(z)$ \cite{Sharov/2018}, and the Pantheon sample \cite{Scolnic/2018}. By using likelihood, we obtain the best fit for the free model parameters.\\
The article organizes as follows: In section \ref{section 2}, we present the field equation formalism in Weyl Type $f(Q,T)$ gravity. In section \ref{section 3}, we describe the cosmological model for the bulk viscous matter-dominated universe and derive the Hubble parameter expression in terms of redshift $z$. In section \ref{section 4}, we constrain the free model parameters $\zeta_{1}$, $\alpha$, and $\beta$ using 57 points of Hubble datasets and 1048 points of Pantheon datasets. We analyze the behavior of the deceleration parameter, the EoS parameter, the energy density, the statefinder diagnostic, and the Om diagnostic in section \ref{section 5}. Finally, in the last section \ref{section 6}, we briefly discuss our results.

\section{Overview of the Weyl type $f(Q,T)$ Gravity} \label{section 2}

The gravitational action in the Weyl-type $f(Q, T)$ is formulated as 
\begin{multline}
\label{1}
 S=\int d^4x\sqrt{-g}\left[ \kappa^2f(Q,T)-\frac{1}{4}W_{\alpha \beta}W^{\alpha \beta}-\frac{1}{2}m^2 w_\alpha w^\alpha+\right.\\ 
 \left. \lambda \tilde{R}+\mathcal{L}_m\right]
\end{multline}    
where, $\tilde{R}=(R+6\nabla_\mu w^\mu-6w_\mu w^\mu)$, $\kappa^2=\frac{1}{16\pi G}$, m is the mass of the particle to the vector field, $\mathcal{L}_m$ is the ordinary matter action. Further, $f$ is an arbitrary function of non-metricity $Q$ and the trace of the matter-energy-momentum tensor $T$. The second and third terms in the action are the ordinary kinetic term and mass term of the vector field respectively.\\ The scalar non-metricity is given by 
\begin{equation}
\label{2}
Q\equiv- g^{\alpha \beta}\left(L^\mu_{\nu\beta}L^\nu_{\beta\mu}-L^\mu_{\nu\mu}L^\nu_{\alpha \beta}\right),
\end{equation}
where, $L^\lambda_{\alpha \beta}$ is the deformation tensor reads as
\begin{equation}
\label{3}
L^\lambda_{\alpha \beta}=-\frac{1}{2}g^{\lambda\gamma}\left(Q_{\alpha\gamma\beta}+Q_{\beta\gamma\alpha}-Q_{\gamma\alpha \beta}\right).
\end{equation}
In Riemannian geometry, the Levi-Civita connection is the compatible with the metric, i.e., $\nabla_\mu g_{\alpha \beta}=0$. This is not the case for the semi-metric connection in Weyl geometry, where we have
\begin{equation}
\label{4}
Q_{\mu\alpha\beta}\equiv\widetilde{\nabla}_\mu g_{\alpha\beta}=\partial_\mu g_{\alpha\beta}-\widetilde{\Gamma}^\rho_{\mu \alpha}g_{\rho\beta}-\widetilde{\Gamma}^\rho_{\mu \beta}g_{\rho\alpha}=2w_\mu g_{\alpha\beta},
\end{equation}
where, $\widetilde{\Gamma}^\lambda_{\alpha\beta}\equiv\Gamma^\lambda_{\alpha\beta}+g_{\alpha\beta}w^\lambda-\delta^\lambda_\alpha w_\beta-\delta^\lambda_\beta w_\alpha$ and $\Gamma^\lambda_{\alpha\beta}$ is the christoffel symbol with respect to the metric $g_{\alpha\beta}$.\\\\
Putting Eq. \eqref{4} in Eq. \eqref{3}, we obtain the relation 
\begin{equation}
\label{5}
Q=-6w^2.
\end{equation}
Varying the action with respect to the vector field, we obtain the generalised Proca equation describing the field evolution,
\begin{equation}\
\label{6}
\nabla^\beta W_{\alpha\beta}-\left(m^2+12\kappa^2f_Q+12\lambda\right)w_\alpha=6\nabla_\alpha \lambda.
\end{equation}
The effective dynamical mass of the vector field is obtained by comparing the above equation with the standard Proca equation.
\begin{equation}
\label{7}
m^2_{eff}=m^2+12\kappa^2f_Q+12\lambda.
\end{equation}
Variation of the action with respect to the metric tensor gives us the following gravitational field equation.
\begin{multline}
\label{8}
\frac{1}{2}\left(T_{\alpha\beta}+S_{\alpha\beta}\right)-\kappa^2f_T\left(T_{\alpha\beta}+\Theta_{\alpha\beta}\right)=-\frac{\kappa^2}{2}g_{\alpha\beta}f\\
-6k^2f_Q w_\alpha w_\beta +\lambda\left(R_{\alpha\beta}-6w_\alpha w_\beta +3g_{\alpha\beta}\nabla_\rho w^\rho \right)\\
+3g_{\alpha\beta}w^\rho \nabla_\rho \lambda 
-6w_{(\alpha}\nabla_{\beta )}\lambda+g_{\alpha\beta}\square \lambda-\nabla_\alpha\nabla_\beta \lambda,
\end{multline}
where, 
\begin{equation}
\label{9}
T_{\alpha\beta}\equiv-\frac{2}{\sqrt{-g}}\frac{\delta(\sqrt{-g}L_m)}{\delta g^{\alpha\beta}},
\end{equation} 
\begin{equation}
\label{10}
f_T\equiv \frac{\partial f(Q,T)}{\partial T},
f_Q\equiv\frac{\partial f(Q,T)}{\partial Q}.
\end{equation}
respectively. Also the expression for $\Theta_{\alpha\beta}$ is defined as
\begin{equation}
\label{11}
\Theta_{\alpha\beta}=g^{\mu\nu}\frac{\delta T_{\mu\nu}}{\delta g_{\alpha\beta}}=g_{\alpha\beta}L_m-2T_{\alpha\beta}-2g^{\mu\nu}\frac{\delta^2 L_m}{\delta g^{\alpha\beta}\delta g^{\mu\nu}}.
\end{equation}
Here, $S_{\alpha\beta}$ is the re-scaled energy momentum tensor of the free Proca field,
\begin{equation}
\label{12}
S_{\alpha\beta}=-\frac{1}{4}g_{\alpha\beta}W_{\rho\sigma}W^{\rho\sigma}+W_{\alpha\rho}W^\rho_\beta -\frac{1}{2}m^2g_{\alpha\beta}w_\rho w^\rho +m^2 w_\alpha w_\beta,
\end{equation}
with $W_{\alpha\beta}=\nabla_\beta w_\alpha-\nabla_\alpha w_\beta$ .\\

\section{Cosmological model for bulk viscous matter dominated universe}
\label{section 3}

We assume that the FLRW metric in the flat space geometry, given by
\begin{equation}
\label{13}
ds^2=-dt^2+a^2(t)\delta_{ij}dx^i dx^j,
\end{equation}
where, $a(t)$ is the scale factor. The vector field is assumed of the form $w_\alpha=\left[\psi (t),0,0,0\right]$ due to spatial symmetry. Hence, $w^2=w_\alpha w^\alpha=-\psi^2(t)$, results in  $Q=-6w^2=6\psi^2(t)$. We also fix the Lagrangian of the perfect fluid to be $\mathcal{L}_m=p$.\\
As a result, the energy momentum tensor for the perfect fluid is given by $T^\alpha_\beta = diag\left(-\rho,p,p,p\right)$ and $\Theta^\alpha_\beta = diag\left(2\rho+p,-p,-p,-p\right)$.

Now, for the cosmological case the flat space constraint, and the generalized Proca equation can be written as

\begin{equation}
\label{14}
\dot{\lambda}=\left(-\frac{1}{6}m^2-2\kappa^2f_Q-2\lambda\right)\psi=-\frac{1}{6}m^2_{eff}\psi ,
\end{equation}

\begin{equation}
\label{15}
\dot{\psi}=\dot{H}+2H^2+\psi^2-3H\psi,
\end{equation} 

\begin{equation}
\label{16}
\partial_i \lambda=0.
\end{equation}

The generalized Friedmann equations is obtained from Eq. \eqref{8} as
\begin{multline}
\label{17}
\kappa^2f_T\left(\rho+\tilde{p}\right)+\frac{1}{2}\rho=\frac{\kappa^2}{2}f-\left(6\kappa^2f_Q+\frac{1}{4}m^2\right)\psi^2 \\
-3\lambda\left(\psi^2-H^2\right)-3\dot{\lambda}\left(\psi-H\right),
\end{multline}

\begin{multline}
\label{18}
-\frac{1}{2}\tilde{p}=\frac{\kappa^2}{2}f+\frac{m^2\psi^2}{4}+\lambda\left(3\psi^2+3H^2+2\dot{H}\right)\\
+\left(3\psi+2H\right)\dot{\lambda}+\ddot{\lambda}.
\end{multline}

Here, we assumed the effective pressure $\bar{p}=p-3H\zeta$, $\zeta$ is the bulk viscosity coefficient. Also, dot $(\cdot)$ represents the derivative with respect to time and $f_Q$ and $f_T$ are the derivatives of $f$ with respect to $Q$ and $T$ respectively.\\
We consider the bulk viscosity coefficient of the form 
\begin{equation}
\label{19}
\zeta=\zeta_0+\zeta_1H+\zeta_2\left(\frac{\dot{H}}{H}+H\right),
\end{equation}
where $\zeta_0$, $\zeta_1$ and $\zeta_2$ are constants and $H$ is Hubble parameter. In this case, the viscosity is related to velocity $\dot{a}$, which is related to the Hubble parameter, and the acceleration $\ddot{a}$. \\
It is assumed that the cold dark matter is highly non-relativistic. So, considering the effect of dark energy on the evolution of the universe through viscous component, which has a dimension of pressure.
Therefore, we can take $p=0$ in this case. Using the functional form,  $f(Q,T)=\alpha Q+\frac{\beta}{6\kappa^2}T$ and effective pressure $\bar{p}$ in Eq. \eqref{17} and \eqref{18}, we obtain the following differential equation. It is worth mentioning that $\beta=0$ corresponds to the $f(Q,T)= \alpha Q$ i.e., a case of the successful theory of General Relativity (GR). Also,  $T=0$, the case of vacuum, the theory reduces to $f(Q)$ gravity and their linear form, i.e.$f(Q)=\alpha Q$ which is equivalent to GR, that passes all Solar System tests.
\begin{equation}
\label{20}
A H^2-B H+C \dot{H}=0,
\end{equation}
where
\begin{multline}
\label{21}
 A=-\left(\frac{9 (\beta +2)^2}{4 \beta }-\frac{\beta}{4}\right) \left(\zeta _1+\zeta _2\right)+\left(\frac{18}{\beta }+6\right)\\
 +36 \left((\alpha +1) \left(\frac{18}{\beta }+12\right)+\left(\frac{3}{2 \beta }+1\right) M^2\right),
\end{multline}

\begin{equation}
\label{22}
B=\left(\frac{9 (\beta +2)^2}{4 \beta }-\frac{\beta}{4}\right) \zeta_0,
\end{equation}
and
\begin{equation}
\label{23}
C=-\left(\left(\frac{9 (\beta +2)^2}{4 \beta }-\frac{\beta}{4}\right) \zeta_2\right)+\frac{12}{\beta }+6,
\end{equation}
We consider $\frac{dH}{dt}=H\frac{dH}{dlna}$ and $a=\frac{1}{1+z}$. Taking $a_{0}=1$, we obtain the required solution of the above differential equation.
\begin{equation}
\label{24}
H\left(z\right)=H_0 (z+1)^{A/C} + \frac{B \left(1-(z+1)^{A/C}\right)}{A}.
\end{equation}
%The expression for the deceleration parameter is 
%\begin{equation}
%\label{25}
%q\left(z\right)=\frac{A H_0-B}{C \left(\frac{B \left((z+1)^{-\frac{A}{C}}-1\right)}{A}+H_0\right)}-1.
%\end{equation}

\section{Observational data analysis} \label{section 4}

The cosmological data used in this study is described here. We use three different current observational datasets to constrain the considered model with a focus on the evidence relevant to the expansion history of the universe, such as the distance-redshift relationship. More crucially, recent research has looked into the role of $H(z)$ and $SNeIa$ data in cosmological constraints and discovered that they could both constrain cosmological parameters. 
In this model, the model parameters are $\zeta_{0}$, $\zeta_{1}$, $\zeta_{2}$, $\alpha$, and $\beta$. In view of large number of free parameters in our model, and in order to do the observational analysis, we fix  $\zeta_0=3.9$ and $\zeta_2=1.415$. Further, we obtained the best fit values of the model parameters  $\zeta_1$, $\alpha$, and $\beta$, by using 57 points from the Hubble dataset, 1048 points from the Pantheon sample and further the combination of Hz+BAO+Pantheon.

\subsection{Hubble data}

\begin{figure*}[htbp]
\centering
\includegraphics[scale=0.5]{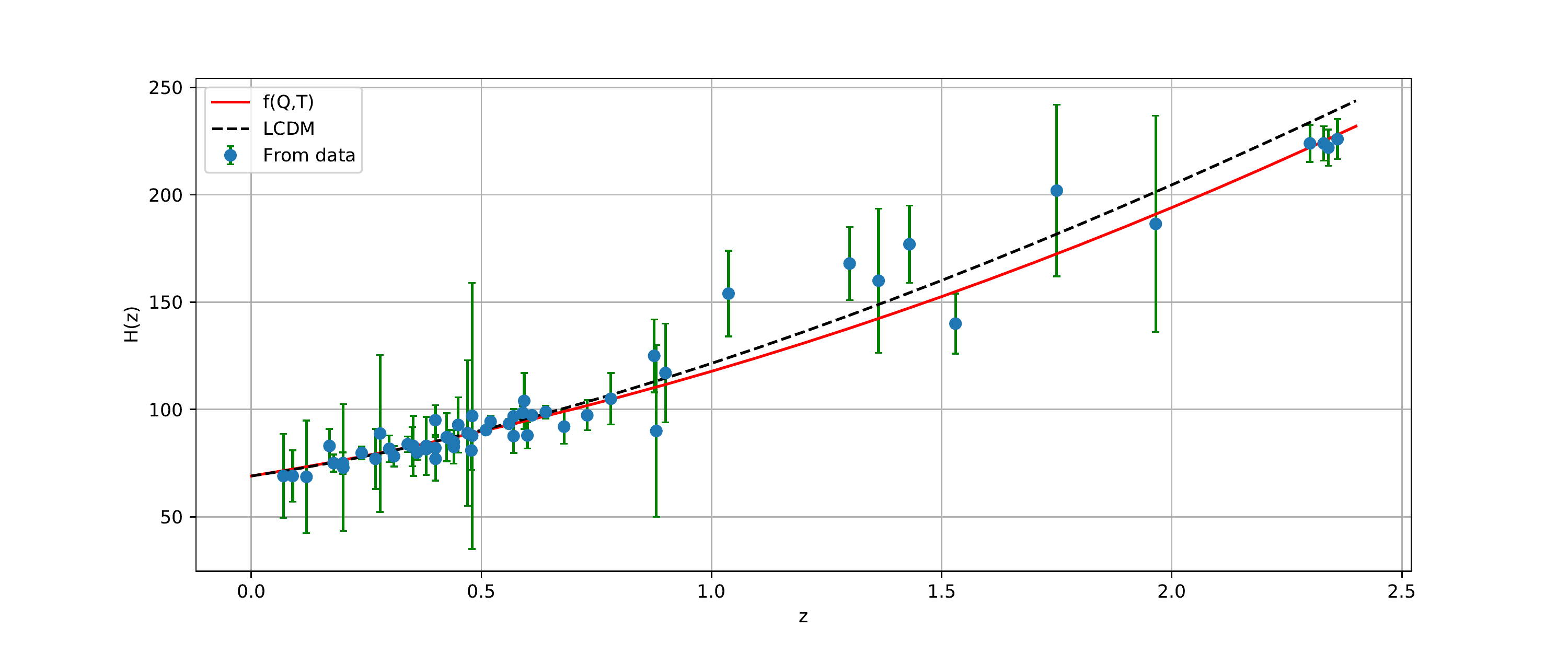}
\caption{The evolution of the $H(z)$ function versus $z$. The red curve shown in the plot is the obtained model. The green dots shown are the Hubble datasets consisting of 57 data points with their corresponding error bars, and also the black dashed line depicts the $\Lambda$CDM model.}
\label{fig-errorHub}
\end{figure*}

The expansion rate $H(z)$ is stated as $H(z)= -\frac{1}{(1+z)} \frac{dz}{dt}$, where $dz$ is obtained through spectroscopic surveys. The Hubble parameter measurements are obtained from early-type galaxies with a passive evolution by estimating their differential age, line of sight. Here, we employ the revised collection of 57 data points including 31 points from the differential age technique, the left 26 points measured using BAO and various redshift ranges $0.07 < z< 2.42$ \cite{Sharov/2018}. So, one can construct a $\chi^{2}$ estimator as 

\begin{equation}
\label{26}
\chi^{2}_{H}= \sum_{i=1}^{57} \dfrac{\left(H_{th}(z_{i}, \zeta_{1}, \alpha, \beta)- H_{obs}(z_{i})\right)^{2}}{\sigma^{2}_{H}(z_{i})}
\end{equation}
where $H_{obs}$ denotes the observed values, $H_{th}$ represents the theoretical values and $\sigma^{2}_{H}(z_{i})$ represents the observational errors on the measured values. 
We have considered $H_{0}= 69$ km/s/Mpc, $\Omega_{\Lambda_{0}}= 0.7$, and $\Omega_{m_{0}}= 0.3$. Also, fig. \ref{fig-errorHub} depicts the error bar plot of 57 points of $H(z)$ and comparison of the considered model with the well-motivated $\Lambda$CDM model.

\begin{figure*}[htbp]
\centering
\includegraphics[scale=0.5]{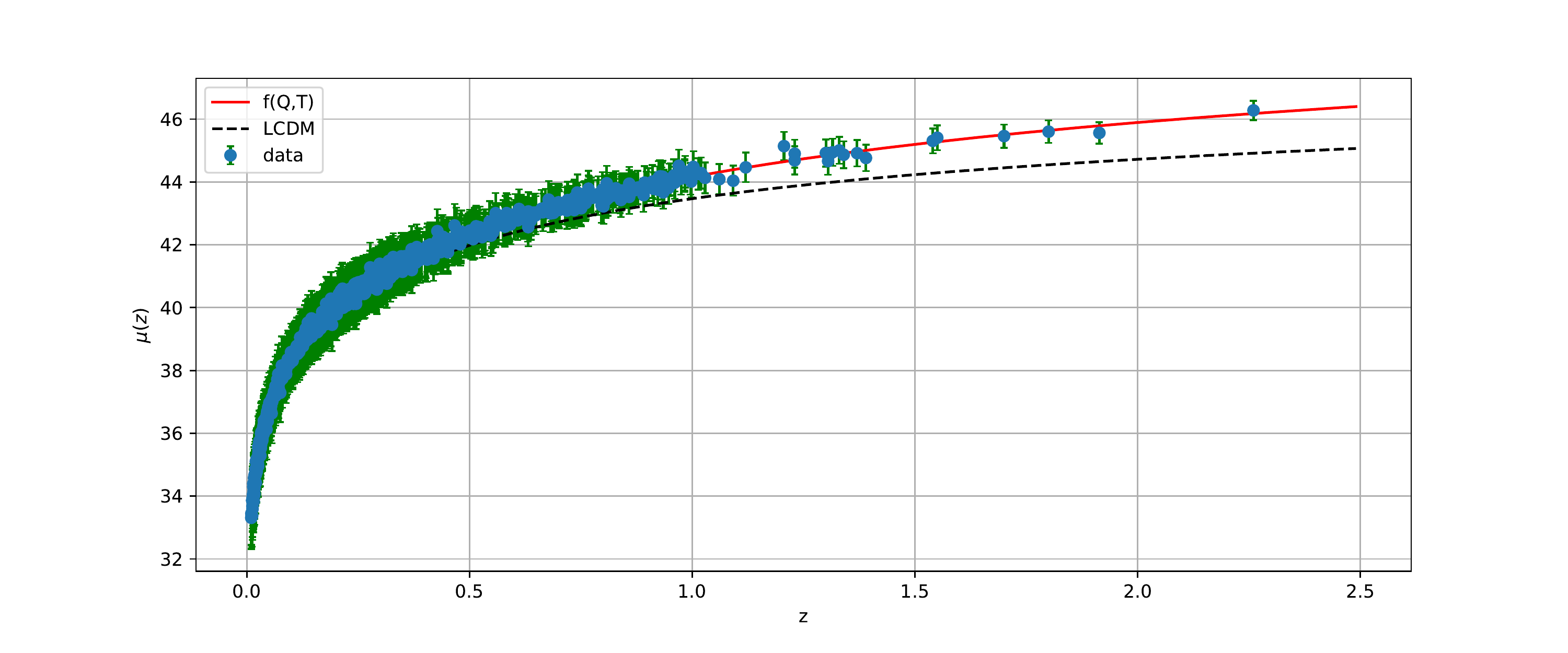}
\caption{The evolution of the $\mu(z)$ function versus $z$. The red curve shown in the plot is the obtained model. The green dots shown are the pantheon samples consisting of 1048 data points and the black dashed line depicts the $\Lambda$CDM model.}
\label{fig-errormuz}
\end{figure*}

\subsection{Pantheon Supernovae data}

\begin{figure*}[htbp]
\centering
\includegraphics[scale=0.7]{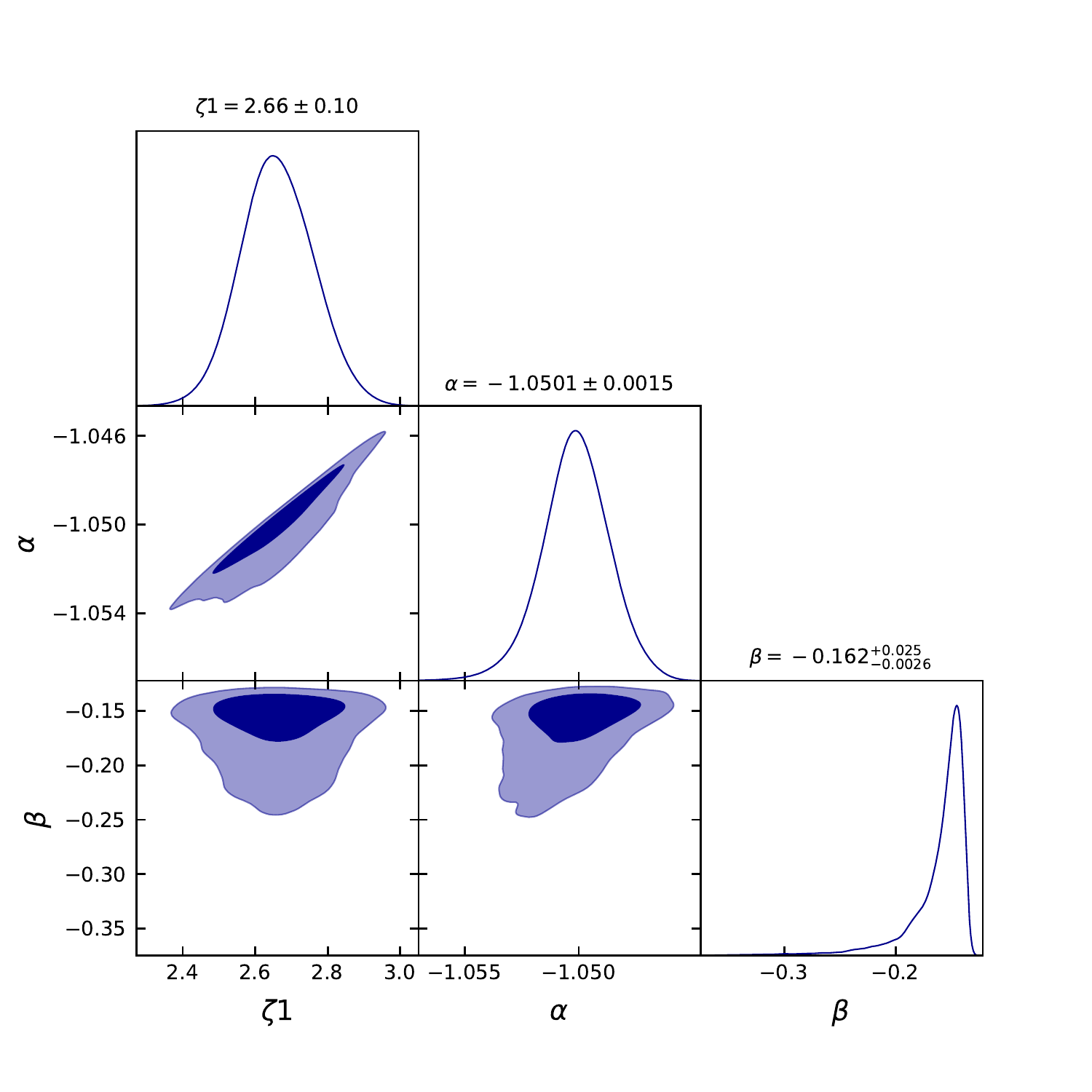}
\caption{ The contour plots for the model parameters  $\zeta_{1}$, $\alpha$, $\beta$ with $1-\sigma$ and $2-\sigma$ confidence limits. It includes the best fit values of the model parameters obtained from the Pantheon samples of 1048 points.}
\label{PantheonC}
\end{figure*}

The pantheon collection is one of the most recent type Ia Supernovae data collections. This set of 1048 SNeIa covering the redshift range $0.01< z<2.26$ \cite{Scolnic/2018,Chang/2019} is chosen and used in the conventional way to establish
\begin{equation} 
\label{27}
\chi^{2}_{P}= \sum_{i=1}^{1048} \dfrac{\left( \mu_{th}(z_{i}, \zeta_{1}, \alpha, \beta)-\mu_{obs}(z_{i})\right) ^{2}}{\sigma^{2}_{H}(z_{i})} 
\end{equation} 
where, $ \mu_{th}$ is the theoretical value of distance modulus read as $\mu_{th}= m-M$ with $m$ as the apparent magnitude for redshift $z_{i}$ and $M$ is a hyperparameter that quantifies uncertainties of various origins. Furthermore, the theoretical distance modulus is defined as $\mu_{th}(z)= 5log\left(\frac{d_{L(z)}}{Mpc}\right)+25$, where the luminosity distance is $d_{L}(z)= c(1+z)\int_{0}^{z} \frac{dz^{'}}{H(z^{'})}$. We calculated the best fit values of the model parameters $\zeta_{1}$, $\alpha$, $\beta$ using pantheon samples that is depicted in fig. \ref{PantheonC} as two dimensional contours with $1-\sigma$ and $2-\sigma$ confidence regions. Moreover, we observed and compared the considered model with the well motivated $\Lambda$CDM model fitting the pantheon sample nicely as shown in fig. \ref{fig-errormuz}.

%\begin{widetext}
\begin{figure*}[htbp]
\centering
\includegraphics[scale=0.7]{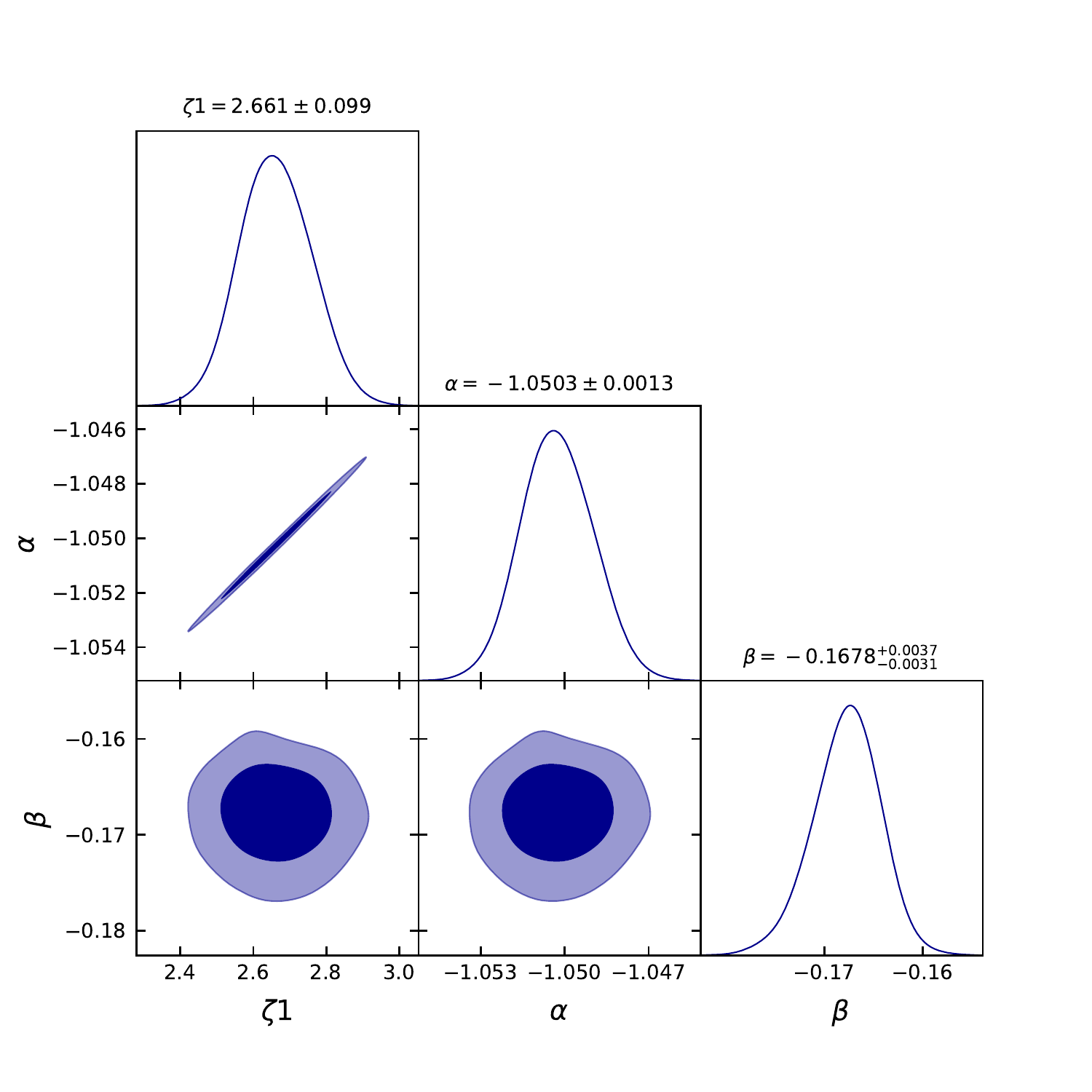}
\caption{ The contour plots for the model parameters  $\zeta_{1}$, $\alpha$, $\beta$ with $1-\sigma$ and $2-\sigma$ confidence limits. It includes the best fit values of the model parameters obtained from the Hubble datasets consisting of 57 points.}
\label{HubbleC}
\end{figure*}
%\end{widetext}

%\begin{widetext}
%\end{widetext}

%\begin{widetext}
\begin{figure*}[htbp]
\centering
\includegraphics[scale=0.7]{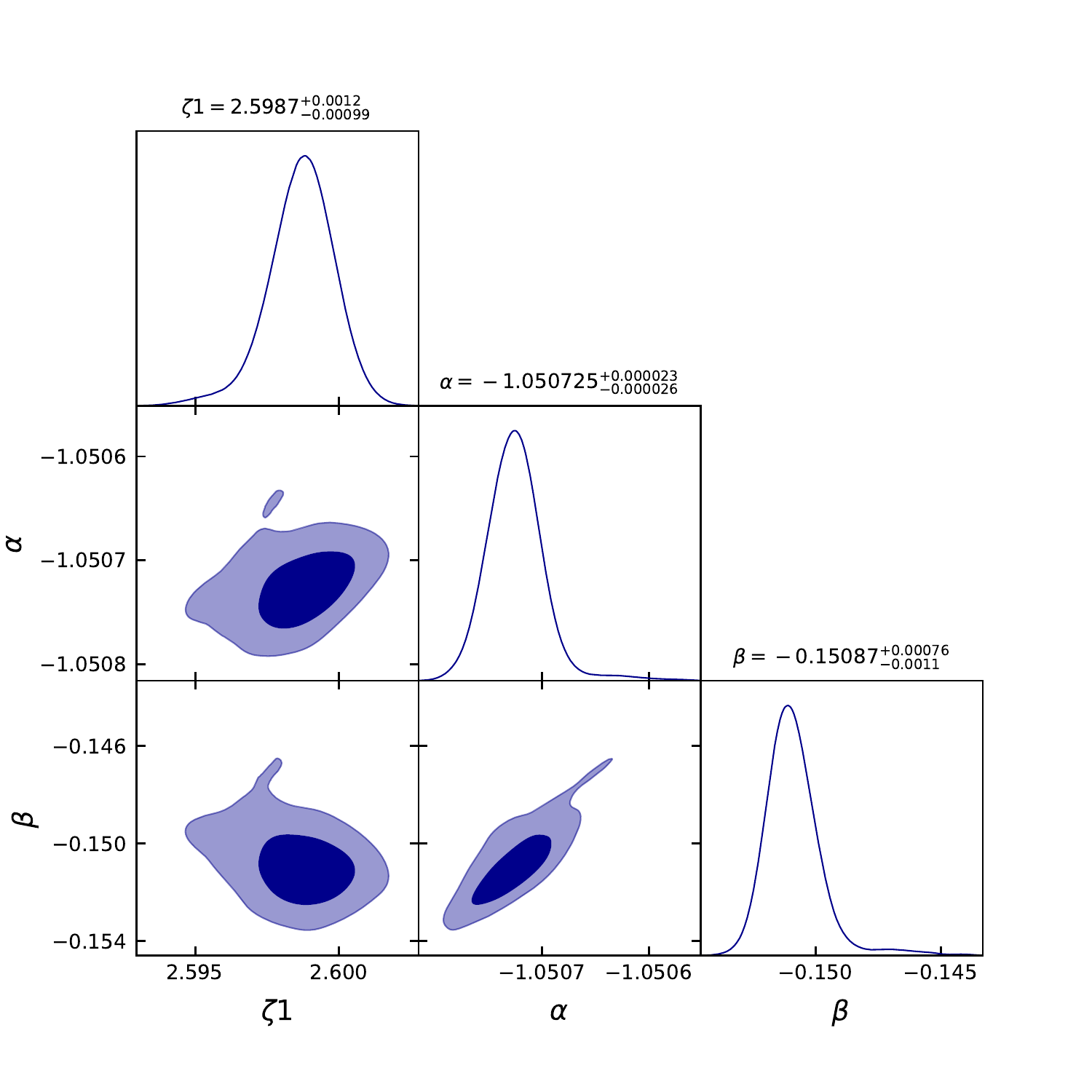}
\caption{ The contour plots for the model parameters  $\zeta_{1}$, $\alpha$, $\beta$ with $1-\sigma$ and $2-\sigma$ confidence limits. It includes the best fit values of the model parameters obtained from the combination Hz+BAO+Pantheon datasets.}
\label{CombineC}
\end{figure*}
%\end{widetext}

\subsection{BAO data}
In the early universe, relativistic sound waves left an impression, which later-time large scale structures could see as baryonic acoustic oscillations (BAO). Here, we take $d_{A}(z^{*})/D_{V}(z_{BAO}))$ \cite{Blake/2011,Percival/2010,Beutler/2011,Jarosik/2011,Eisenstein/2005,Giostri/2012} , where $z^{*}$ is the radiation-matter decoupling time represented by $z^{*}\approx 1091$, $d_{A}$ is the co-moving angular diameter distance given as $d_{A}=\int_{0}^{z} \frac{d\overline{z}}{H(\overline{z})}$, and $D_{V}= \left(d_{A}(z)^{2} \frac{z}{H(z)}\right)^{1/3}$ is the dilation scale. The chi square function for BAO is used as
\begin{equation}
\chi^{2}_{BAO}= X^{T} C^{-1} X,
\end{equation}

where 
$$
\begin{bmatrix}
\frac{d_{A}(z^{*})}{D_{V}(0.106)}-30.95\\
\frac{d_{A}(z^{*})}{D_{V}(0.2)}-17.55\\
\frac{d_{A}(z^{*})}{D_{V}(0.35)}-10.11\\
\frac{d_{A}(z^{*})}{D_{V}(0.44)}-8.44\\
\frac{d_{A}(z^{*})}{D_{V}(0.6)}-6.69\\
\frac{d_{A}(z^{*})}{D_{V}(0.73)}-5.45
\end{bmatrix}
$$
and the inverse covariance matrix $C^{-1}$ is defined in \cite{Giostri/2012}

\subsection{Results}
The likelihood contours for our model parameters, with $1-\sigma$ and $2-\sigma$, are shown in the figures \ref{HubbleC}, \ref{PantheonC} and \ref{CombineC} using the above data samples. We minimize the chi-square for Hubble and Pantheon independently and finally with the combination for $Hz+BAO+Pantheon$. The constrained values of the model parameters are summarized in Table 1.

\begin{table*}[htbp]
\centering
\begin{tabular}{ |c|c|c|c| }
 \hline
 \multicolumn{4}{|c|}{Best fit values} \\
 \hline
 Parameters & Hubble data & Pantheon data & Hz+BAO+Pantheon \\
 \hline
 $\zeta_{1}$   & $2.661^{+0.099}_{-0.099}$    & $2.66^{+0.10}_{-0.10}$ & $2.5987^{+0.00120}_{-0.00099}$\\
 \hline
$\alpha$ &  $-1.0503^{+0.0013}_{-0.0013}$  & $-1.0501^{+0.0015}_{-0.0015}$ & $-1.0507^{+0.000023}_{-0.000026}$ \\
\hline
 $\beta$ & $-0.1678^{+0.0037}_{-0.0031}$ & $-0.162^{+0.0250}_{-0.0026}$ & $-0.1508^{+0.00076}_{-0.00110}$ \\
 \hline
% \multicolumn{4}{|c|}{The present value of the deceleration parameter } \\
 %\hline
 $q_0$   & $-0.54^{+0.1294}_{-0.1890}$ & $-0.42^{+0.2363}_{-1.9100}$ & $-0.59^{+0.0267}_{-0.1063}$ \\
 \hline
 %\multicolumn{4}{|c|}{The value of transition redshift } \\
 %\hline
 $z_t$  & $1.05^{+1.0520}_{-0.4256}$ & $0.506^{+0.912}_{-0.025}$ & $0.686^{+0.0300}_{-0.1002}$\\
 \hline
 %\multicolumn{4}{|c|}{The present value of EoS } \\
% \hline
 $w_0$  & $-0.3731^{+0.1038}_{-0.3111}$ & $-0.2699^{+0.0793}_{-0.2115}$ & $-0.3804^{+0.0236}_{-0.1347}$\\
 \hline
 %\multicolumn{4}{|c|}{The present value of $(s,r)$ parameters} \\
 %\hline
 $s_0$  & $0.1071^{+0.0107}_{-0.0343}$ & $0.0659^{+0.0204}_{-0.3095}$ & $0.0228^{+0.0103}_{-0.0063}$\\
 \hline
$r_0$ & $0.6643^{+0.0663}_{-0.0138}$  & $0.8186^{+0.528}_{-0.118}$ & $0.925^{+0.0357}_{-0.0294}$\\
\hline
\end{tabular}
\end{table*}

\section{Cosmological parameters} \label{section 5}

\subsection{Deceleration parameter}

According to cosmological observations, the apparent cosmic acceleration is likewise a recent phenomenon. In the absence of DE or when its effect is minor, the identical model should have deceleration in the early period of the matter era to allow structure production. As a result, to describe the entire evolutionary history of the universe, a cosmological model requires both a decelerated and an accelerated phase of expansion. Hence, it is significant to study the behavior of the deceleration parameter which is defined as

\begin{equation}
\label{28}
q=-\frac{\dot{H}}{H^2}-1
\end{equation}

\begin{figure}[H]
\includegraphics[scale=0.6]{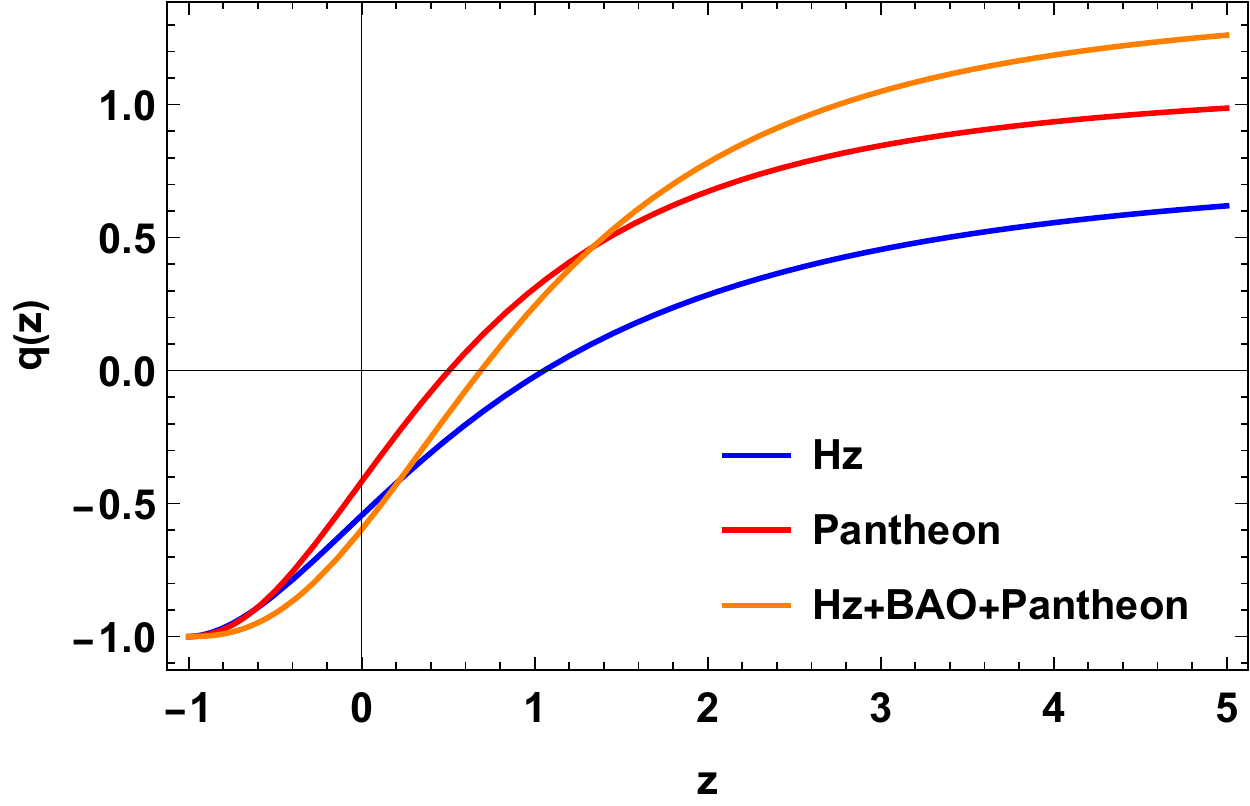}
\caption{The figure shows the trajectory of deceleration parameter $q$ versus redshift $z$ corresponding to the values of the parameters $\zeta_{1}=2.661$, $\alpha=-1.0503$, $\beta=-0.1678$ obtained by Hubble datasets, $\zeta_{1}=2.66$, $\alpha=-1.0501$, $\beta=-0.162$ obtained by Pantheon datasets and $\zeta_{1}=2.5987$, $\alpha=-1.0507$, $\beta=-0.1508$ obtained by Hz+BAO+Pantheon datasets. }
\label{dec}
\end{figure}

We shall write the deceleration parameter $q$ in terms of cosmological redshift $z$ using the relation
(assuming $a(0)=1$)
\begin{equation}
\label{29}
1+z=\frac{1}{a}
\end{equation} 
The deceleration parameter $q$ is obtained in terms of redshift $z$ as

\begin{equation}
\label{30}
q\left(z\right)=\left(1+z\right)\frac{1}{H(z)}\frac{dH(z)}{dz}-1
\end{equation} 
Putting Eq.\eqref{24} in Eq.\eqref{30}, we obtained the deceleration parameter $q$ for our model.
\begin{equation}
\label{31}
q\left(z\right)=\frac{A H_0-B}{C \left(\frac{B \left((z+1)^{-\frac{A}{C}}-1\right)}{A}+H_0\right)}-1,
\end{equation}
where $A$, $B$ and $C$  are constants given in Eqs.\eqref{21}-\eqref{23}. 
Fig. \ref{dec} shows the behavior of deceleration parameter $q$ for the corresponding values of model parameters constrained by the Hubble datasets, Pantheon datasets and Hz+BAO+Pantheon datasets. It indicates that our model successfully generates the late-time cosmic acceleration and the deceleration expansion in the past.  The present value of the deceleration parameter corresponding to the Hubble, Pantheon and Hz+BAO+Pantheon datasets is  $q_0=-0.54^{+0.1294}_{-0.1890}$, $q_0=-0.42^{+0.2363}_{-1.9100}$ and $q_0=-0.59^{+0.0267}_{-0.1063}$ \cite{Santos/2016}, respectively. 

\subsection{Equation of state parameter(EoS)}
\begin{figure}[H]
\includegraphics[scale=0.65]{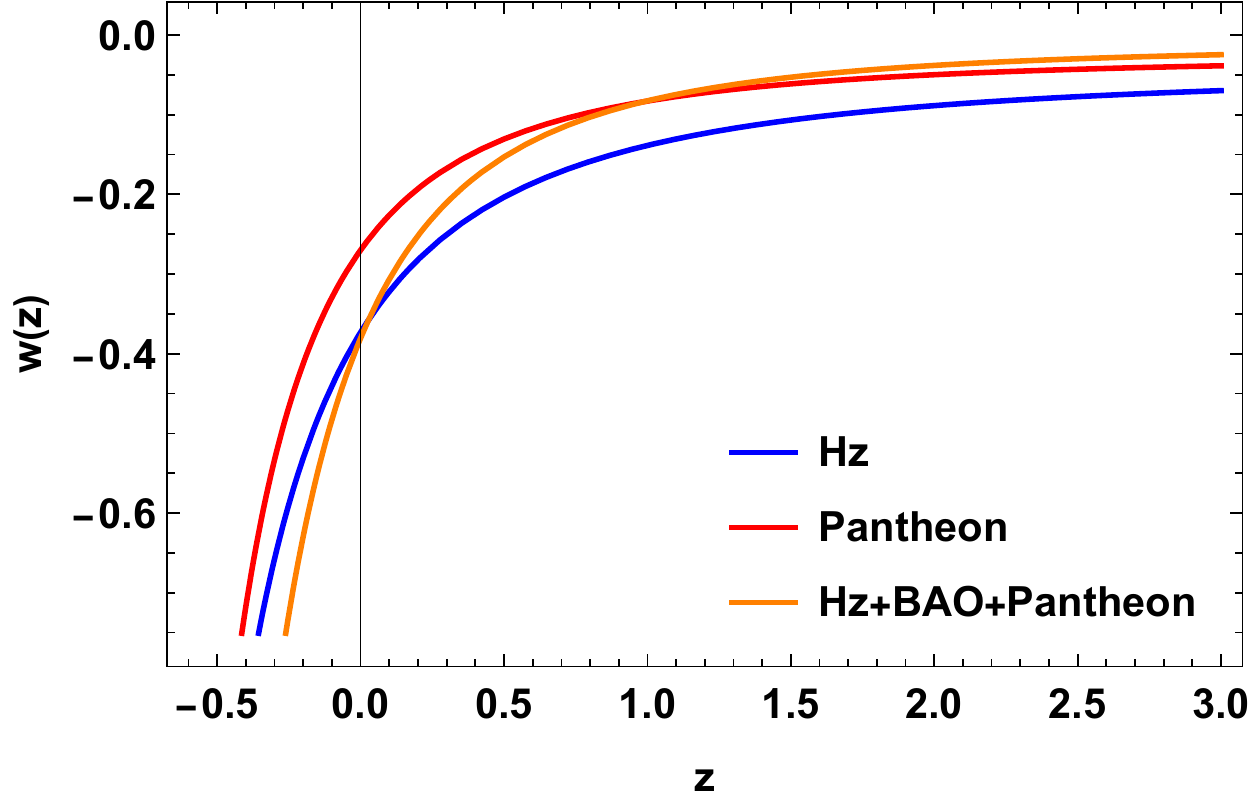}
\caption{The figure shows the trajectory of EoS parameter $w$ versus redshift $z$ corresponding to the values of the parameters $\zeta_{1}=2.661$, $\alpha=-1.0503$, $\beta=-0.1678$ obtained by Hubble datasets, $\zeta_{1}=2.66$, $\alpha=-1.0501$, $\beta=-0.162$ obtained by Pantheon datasets and $\zeta_{1}=2.5987$, $\alpha=-1.0507$, $\beta=-0.1508$ obtained by Hz+BAO+Pantheon datasets.}
\label{w}
\end{figure}

The EoS parameter is defined as the pressure $p$ to energy density $\rho$ ratio, i.e.$w=\frac{p}{\rho}$. The EoS of dark energy can characterize the cosmic inflation and accelerated expansion of the universe. The condition for an accelerating universe is $w<-\frac{1}{3}$. In the simple case, $w=-1$ corresponds to the cosmological constant, i.e,$\Lambda$CDM. Also, the value of EoS $w=\frac{1}{3}$ and $w=0$ shows the radiation and matter-dominated universe, respectively.\\
Figure \ref{w}, shows the behavior of the equation of state parameter. The present value of the EoS parameter is $w_0=-0.3731^{+0.1038}_{-0.3111}$, $w_0=-0.2699^{+0.0793}_{-0.2115}$ and $w_0=-0.3804^{+0.0236}_{-0.1347}$ corresponding to the parameters constrained by the Hubble dataset, Pantheon dataset, and Hz+BAO+Pantheon, respectively. These values clearly show that the present universe is an accelerating phase and lies in the quintessence phase.

\subsection{Density parameter}

\begin{figure}[H]
\includegraphics[scale=0.6]{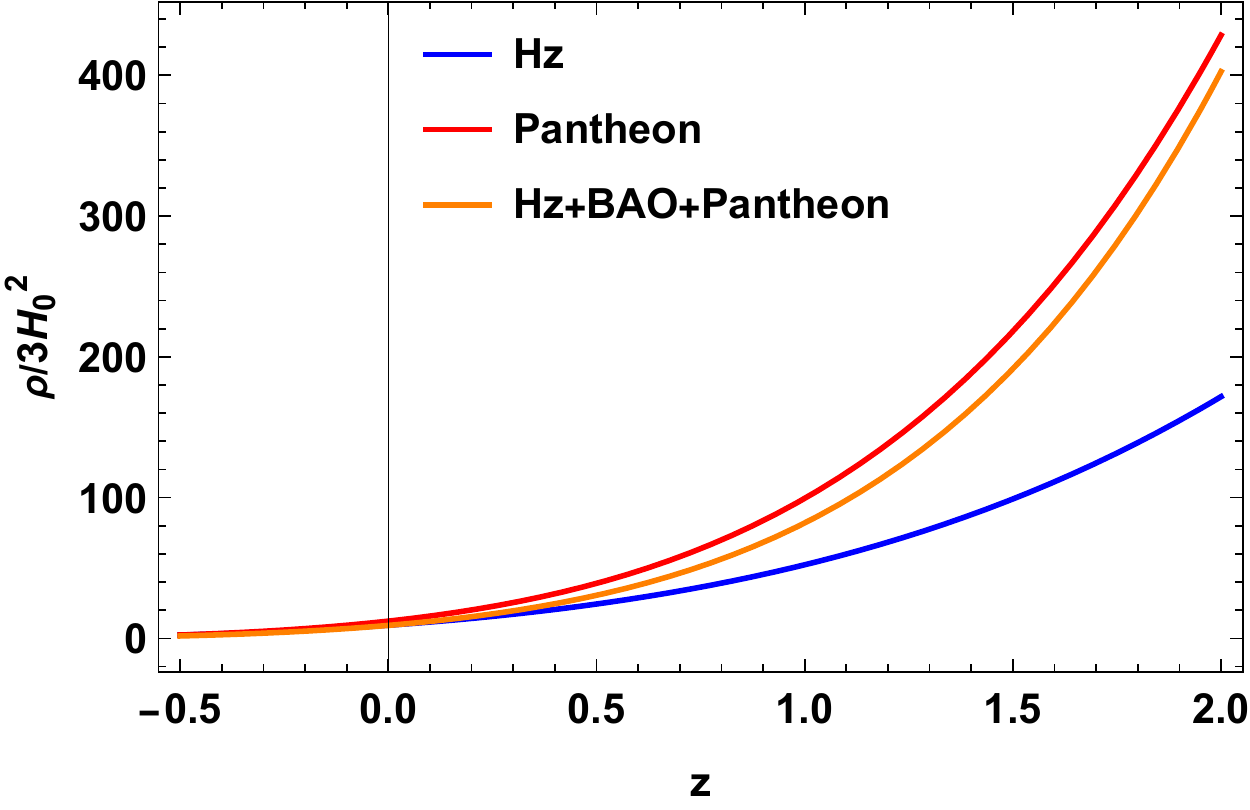}
\caption{The figure shows the trajectory of energy density $\frac{\rho}{3H_{0}^{2}}$ versus redshift $z$ corresponding to the values of the parameters $\zeta_{1}=2.661$, $\alpha=-1.0503$, $\beta=-0.1678$ obtained by Hubble datasets, $\zeta_{1}=2.660$, $\alpha=-1.0501$, $\beta=-0.162$ obtained by Pantheon datasets and $\zeta_{1}=2.5987$, $\alpha=-1.0507$, $\beta=-0.1508$ obtained by Hz+BAO+Pantheon datasets.}
\label{density}
\end{figure}

The density parameter $\rho$ is obtained in terms of $z$ by solving Eq.\eqref{17}, \eqref{18} and using Eq.\eqref{24}. We plot the density parameter for the values of model parameters constrained by the Hubble, Pantheon  and Hz+BAO+Pantheon datasets. It is observed in figure \ref{density}, that the density parameter shows a positive behavior with redshift $z$. Further, the density of the bulk viscous matter-dominated universe falls off with the expansion of the universe in the far future.  

\subsection{Statefinder diagnostics}

As more DE models are developed and developing for understanding cosmic acceleration, it becomes more difficult to distinguish between the different DE models. A sensitive and reliable diagnostic for DE models is essential to differentiate between different cosmological scenarios. 
V. Sahni at el. \cite{Sahni/2003} introduced a new geometrical parameter pair $(s,r)$ and $(q,r)$ called statefinder diagnostics, which recognizes the various dark energy models (DE) for interpreting the cosmic acceleration of the universe \cite{Sahni/2003,Alam/2003,Sami/2012,Rani/2015}. The statefinder analyzes the expansion dynamics of the universe using higher derivatives of the scale factor $\dddot{a}$ and is a suitable associate to the deceleration parameter, which is based on $\ddot{a}$. The $\left\lbrace s,r\right\rbrace $ statefinder pair is defined as

\begin{align}
r &=\frac{\dddot{a}}{a H^3},\\
s&=\frac{r-1}{3(q-\frac{1}{2})}.
\end{align}

\begin{figure}[H]
\includegraphics[scale=0.33]{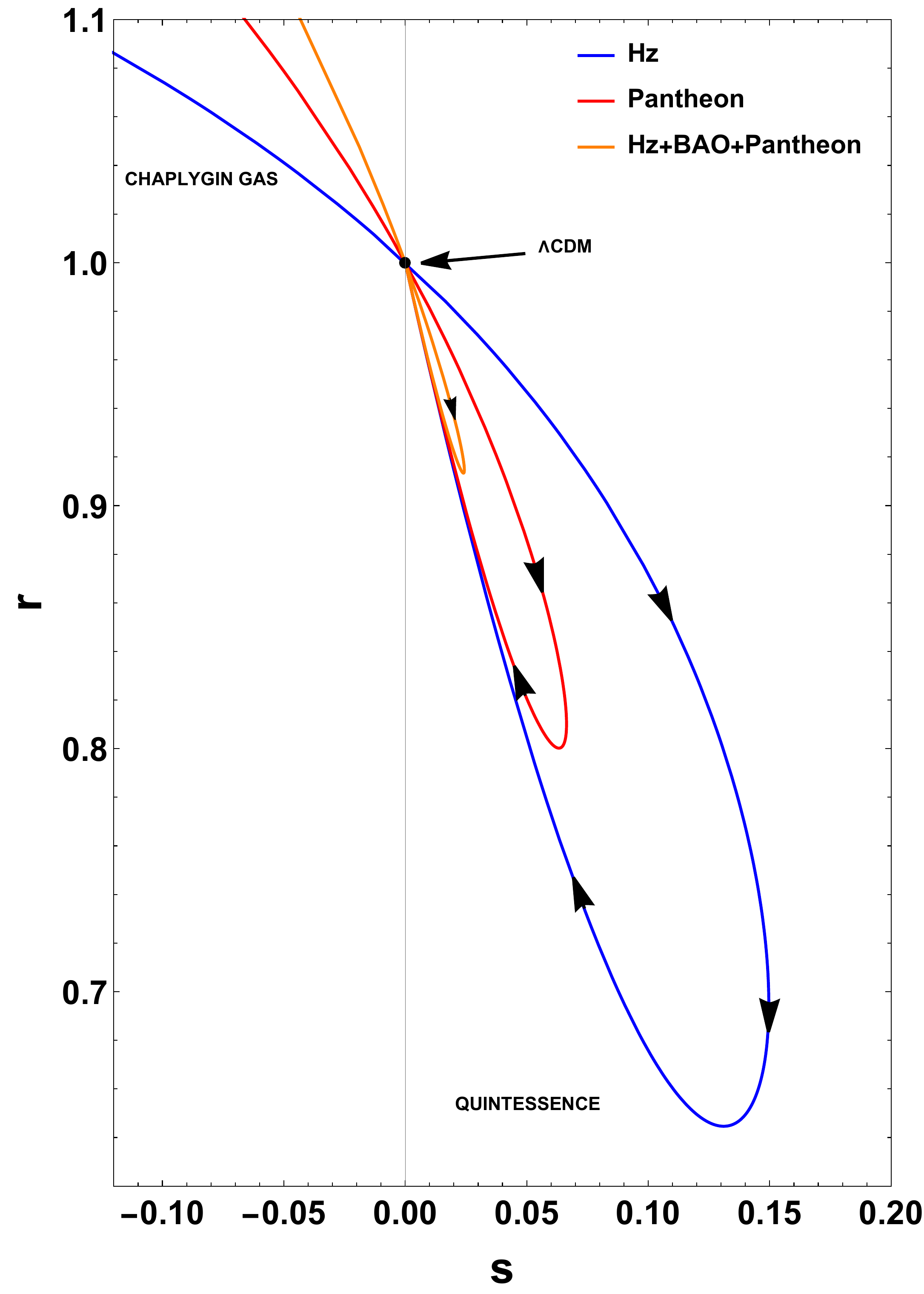}
\caption{The figure shows that the behavior of the statefinder $s-r$ plane corresponding to the values of the model parameters $\zeta_{1}=2.661$, $\alpha=-1.0503$, $\beta=-0.1678$ obtained by Hubble datasets, $\zeta_{1}=2.66$, $\alpha=-1.0501$, $\beta=-0.162$ obtained by Pantheon datasets and $\zeta_{1}=2.5987$, $\alpha=-1.0507$, $\beta=-0.1508$ obtained by Hz+BAO+Pantheon datasets. }
\label{rs}
\end{figure}

\begin{figure}[H]
\includegraphics[scale=0.4]{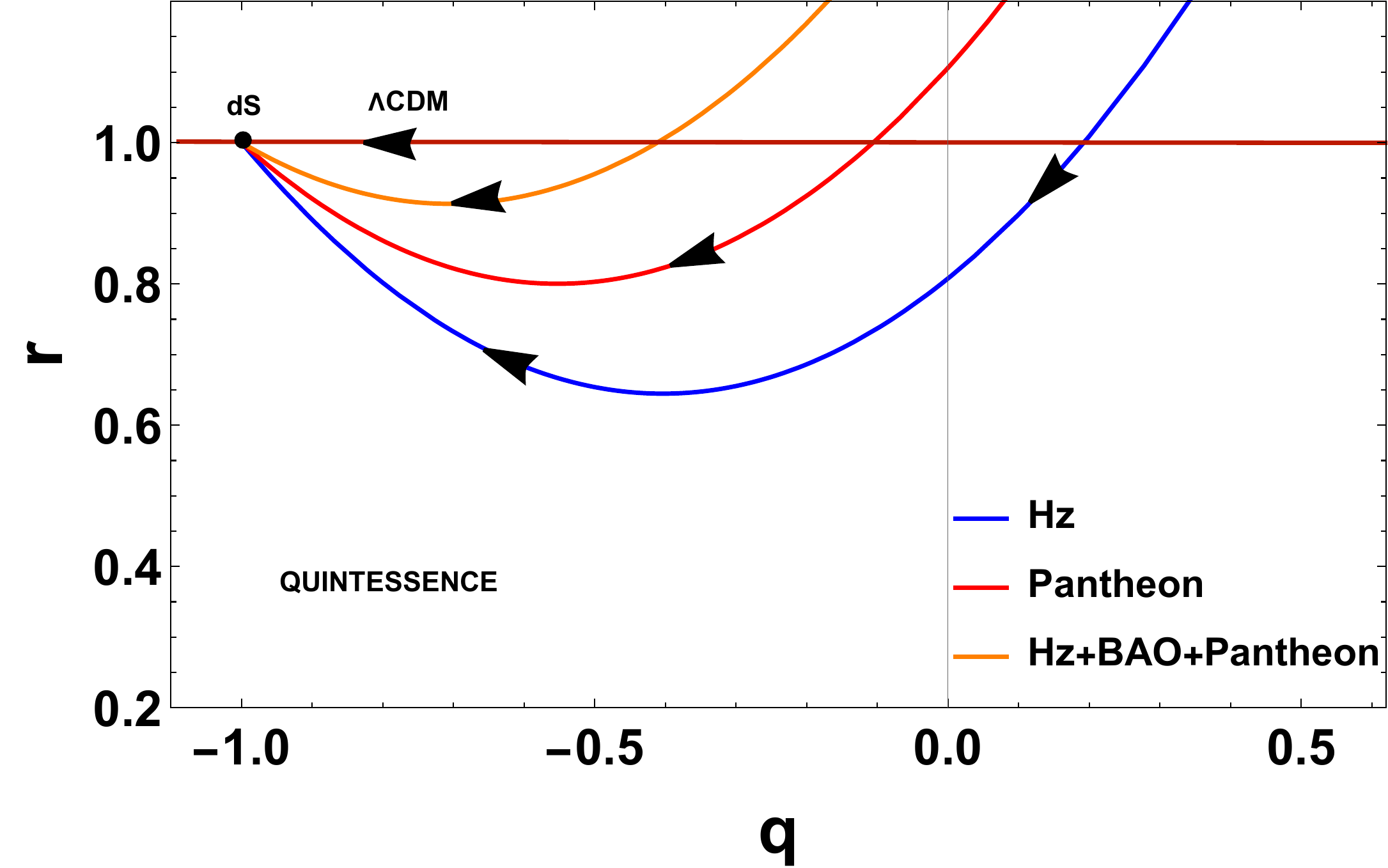}
\caption{The plot shows that the behavior of the statefinder $q-r$ plane corresponding to the values of the model parameters $\zeta_{1}=2.661$, $\alpha=-1.0503$, $\beta=-0.1678$ obtained by Hubble datasets, $\zeta_{1}=2.66$, $\alpha=-1.0501$, $\beta=-0.162$ obtained by Pantheon datasets and $\zeta_{1}=2.5987$, $\alpha=-1.0507$, $\beta=-0.1508$ obtained by Hz+BAO+Pantheon datasets.}
\label{rq}
\end{figure}

The evolution trajectories of statefinder pair $\left\lbrace s,r\right\rbrace $ and $\left\lbrace q,r\right\rbrace $ are shown in figures \ref{rs} and \ref{rq}, respectively using the constrained values of model parameters. The trajectory is primarily contained in the first quadrant of the s-r plane. In figure \ref{rs}, the fixed point $(0,1)$ corresponds to the spatially flat $\Lambda$CDM model. Thus, the distance between a particular cosmological model and the $\Lambda$CDM model in s-r plane, such as quintessence, Chaplygin gas, phantom and interactive dark energy models can be easily outlined as examined in the literature \cite{Alam/2003,Cao/2018,Wu/2010}. The trajectory of $(s,r)$ pair comes from the Chaplygin gas type dark energy model and lies in the Quintessence region i.e., $s>0$ and $r<1$. The present values of (s,r) parameters are (0.1071, 0.6643), (0.0659, 0.8186) and (0.0228,0.925) corresponding to the parameters constrained by the Hubble, Pantheon and Hz+BAO+Pantheon datasets, respectively \cite{Solanki/2021}. Furthermore, all evolutionary trajectories move away from the $\Lambda$CDM until they revert to it. \\
Figure \ref{rq} shows that the model evolves from the point $(q,r)=(0.5,1)$ in the past, which corresponds to a matter dominated $SCDM$ universe and end the evolution at $(q,r)=(-1,1)$, the de Sitter (dS) point. Eventually, both tend to evolve like a $\Lambda$CDM universe/de Sitter point, i.e., $\{r, s\} = \{1, 0\}$, or $\{q, r\} = \{-1, 1\}$ in the future.

\subsection{Om Diagnostics} 
\begin{figure}[H]
\includegraphics[scale=0.66]{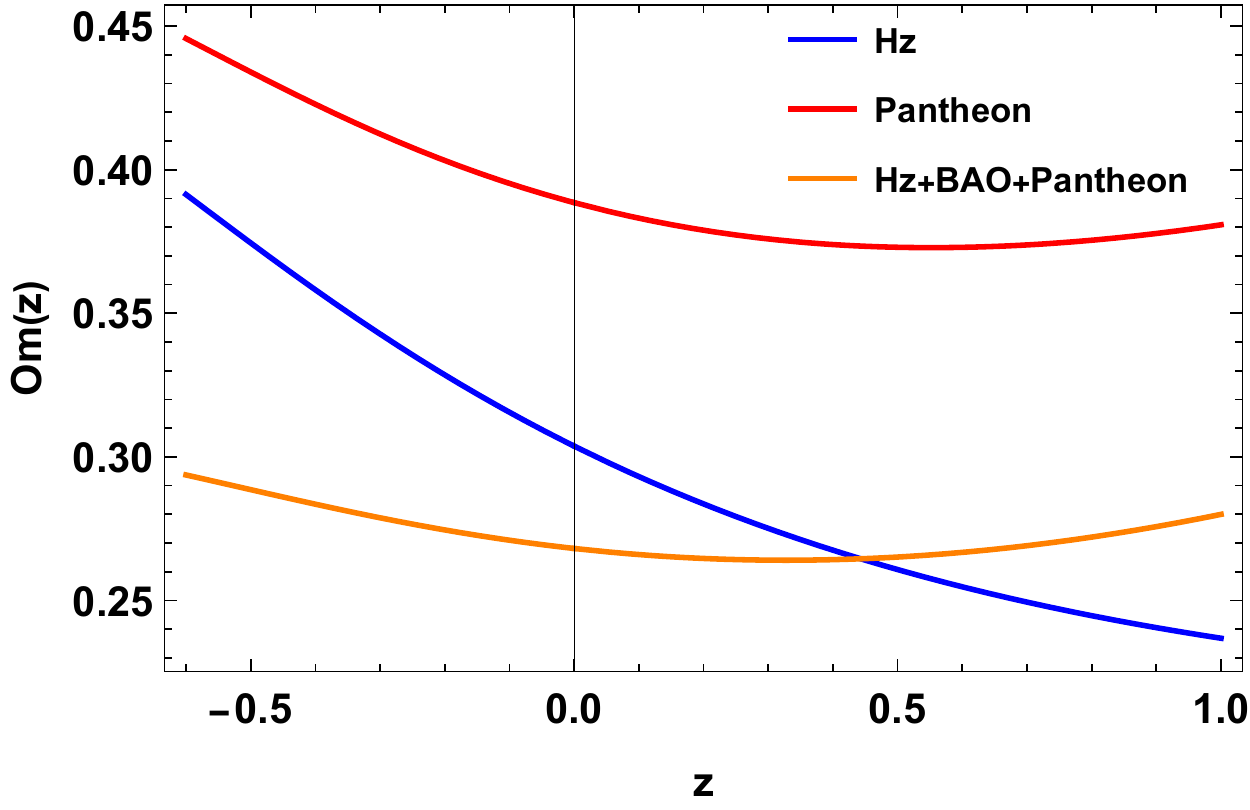}
\caption{The plot shows that the behavior of the $Om$ versus redshift $z$ plane corresponding to the values of the model parameters $\zeta_{1}=2.661$, $\alpha=-1.0503$, $\beta=-0.1678$ obtained by Hubble datasets, $\zeta_{1}=2.66$, $\alpha=-1.0501$, $\beta=-0.162$ obtained by Pantheon datasets and $\zeta_{1}=2.5987$, $\alpha=-1.0507$, $\beta=-0.1508$ obtained by Hz+BAO+Pantheon datasets.}
\label{Om}
\end{figure}

As a complement to the statefinder diagnostic, Sahni et al. \cite{Sahni/2008} proposed a new diagnostic called $Om(z)$ diagnostic, which is a combination of  the Hubble parameter $H=\frac{\dot{a}}{a}$ and the cosmological redshift $z$. The $Om(z)$ is solely determined by the first-order time derivative of the expansion factor, i.e., $\dot{a}$. This  diagnostic can be used to discriminate between different dark energy models by monitoring the slope of $Om(z)$.
The $Om(z)$ diagnostic for spatially flat universe is defined as
\begin{equation}
\label{34}
Om\left(z\right)=\frac{\left(\frac{H(z)}{H_0}\right)^2-1}{\left(1+z\right)^3-1},
\end{equation}

where $H_0$ is the Hubble constant. We have a different set of values of $Om(z)$ for the quintessence, phantom, and the $\Lambda$CDM model. The negative slope of $Om(z)$ demonstrates that dark energy (DE) behaves as a quinte-ssence type $(\omega > -1)$, the positive slope of $Om(z)$ shows that the behavior of dark energy as phantom type $(\omega<-1)$, and the constant behavior of $Om(z)$ demonstrates dark energy as the cosmological constant ($\Lambda$CDM) \cite{Wu/2010}. In figure \ref{Om}, we can easily observe the evolution trajectory of $Om(z)$ diagnostic with negative slope. It depicts that our model lies in a quintessence phase.

\section{Conclusions} \label{section 6}

Weyl-type $f(Q, T)$ gravity is a recently proposed modified theory that explains cosmic acceleration without the need for an additional dark energy component. In the present article, we have examined the impact of bulk viscosity on dynamics of the universe in the Weyl-type $f(Q,T)$ gravity, where $Q$ is a non-metricity, and $T$ is the trace of the energy-momentum tensor. We used the simplest $f(Q,T)$ functional form of $f(Q,T)=\alpha Q+\frac{\beta}{6\kappa^2}T$, where $\alpha$, $\beta$ are constants and the bulk viscous coefficient as $\zeta=\zeta_0+\zeta_1H+\zeta_2\left(\frac{\dot{H}}{H}+H\right)$, where $\zeta_0$, $\zeta_1$ and $\zeta_2$ are constants. We found the exact solutions of the field equations of the Weyl-type $f(Q, T)$ gravity in the presence of bulk viscous matter.\\
In view of the large number of free parameters in our model, in order to do the observational analysis, we fixed  $\zeta_0=3.9$ and $\zeta_2=1.415$. Further, we obtained the best fit values of the model parameters $\zeta_1$, $\alpha$, and $\beta$ by using the 57 points Hubble datasets, the 1048 point Pantheon datasets, and the Hz+BAO+Pantheon datasets.\\
Corresponding to these best fit-values, we have investigated the evolution of the different cosmological parameters. We observed the trajectory of the deceleration parameter $q$, which shows the universe transitioning from the decelerating phase ($q$ is positive) to the accelerating phase ($q$ is negative), and the current value deceleration parameter corresponding to the Hubble, Pantheon and Hz+BAO+Pantheon datasets is $q_0=-0.54^{+0.1294}_{-0.1890}$, $q_0=-0.42^{+0.2363}_{-1.9100}$ and $q_0=-0.59^{+0.0267}_{-0.1063}$, respectively. The EoS parameter shows the negative behavior depicting that the present universe is accelerating and lies in the quintessence phase. The density parameter shows positive behavior for both the constrained values of parameters. Moreover, the evolutionary trajectories of statefinder and Om diagnostics show the deviation of our considered bulk viscous model from other DE models. It is clear that our viscous matter-dominated model lies in the quintessence region and converges to the $\Lambda$CDM fixed point. Lastly, we conclude that the viscosity of cosmic matter content plays a considerable role in driving the universe accelerated expansion. As a result, the geometrical, astrophysical, and cosmological consequences of the Weyl-type $f(Q, T)$ gravity can be studied on theoretical grounds.

\section*{Acknowledgments}

GG acknowledges University Grants Commission (UGC), New Delhi, India for awarding Junior Research Fellowship (UGC-Ref. No.: 201610122060). SA acknowledges CSIR, Govt. of India, New Delhi, for awarding Junior Research Fellowship. PKS acknowledges CSIR, New Delhi, India for financial support to carry out the Research project [No.03(1454)/19/EMR-II Dt.02/08/2019]. We are very much grateful to the honorable referee and the editor for the illuminating suggestions that have significantly improved our work in terms of research quality and presentation.

%%%%%%%%%%%%%%%%%%%%%%%%%%%%%%%%%%%%%%%%%%%%%%%%%%%%%%%%%%%%%%%%%%%%%%%%%%%%%%%

\end{document}